\documentclass[sigconf]{acmart}

\usepackage{amsmath}   
\usepackage{multirow}  
\usepackage{enumitem}  
\usepackage{listings}  

\lstdefinestyle{promptstyle}{
  basicstyle=\ttfamily\scriptsize,
  breaklines=true,
  breakatwhitespace=false,
  columns=fullflexible,
  keepspaces=true,
  frame=single,
  framesep=4pt,
  showstringspaces=false,
  postbreak=\mbox{\textcolor{gray}{$\hookrightarrow$}\space},
  aboveskip=4pt,
  belowskip=4pt
}

\graphicspath{{./}}

\newcommand{\acc}[3]{\shortstack{#1\\[1pt]{\scriptsize[#2,\,#3]}}}

\newlist{rqlist}{description}{1}
\setlist[rqlist]{
  font=\normalfont\bfseries,
  labelwidth=1cm,
  labelsep=0.3em,
  leftmargin=1.3cm,
  itemsep=2pt,
  parsep=0pt,
  topsep=4pt
}

\AtBeginDocument{%
  }

\setcopyright{rightsretained}
\copyrightyear{2026}
\acmYear{2026}
\makeatletter\@printpermissionfalse\makeatother
\acmConference{KDD 2026 Workshop on Evaluation and Trustworthiness of Agentic AI}{August 9, 2026}{Jeju, Republic of Korea}
\acmISBN{}
\acmDOI{}

\begin{document}

\title[IFCMemoryBench]{IFCMemoryBench: Evaluating Long-Term Memory of
  LLM-Based Agents in BIM Information Retrieval}

\author{Changyu Du}
\affiliation{%
  \institution{Technical University of Munich}
  \city{Munich}
  \country{Germany}}
\email{changyu.du@tum.de}

\author{Alexander Vosseler}
\affiliation{%
  \institution{Nemetschek Group}
  \city{Munich}
  \country{Germany}}
\email{avosseler@nemetschek.com}

\author{Filippo Mazza}
\affiliation{%
  \institution{Nemetschek Group}
  \city{Munich}
  \country{Germany}}
\email{fmazza@nemetschek.com}

\author{André Borrmann}
\affiliation{%
  \institution{Technical University of Munich}
  \city{Munich}
  \country{Germany}}
\email{andre.borrmann@tum.de}

\renewcommand{\shortauthors}{Du et al.}

\begin{abstract}
  Long-term memory is becoming a core capability of LLM-based agents, but existing evaluations largely test conversational recall in open-domain or persona-grounded settings. We argue that a stronger test is whether an agent can reuse information from prior sessions while acting over a live, structured, domain-specific environment. We study this problem in Building Information Modelling (BIM), a professional engineering workflow where agents must query large IFC models while also relying on project specifications, client decisions, and engineering conventions that are often discussed in conversation but absent from the model.
We introduce IFCMemoryBench, a human-validated benchmark for evaluating long-term memory in LLM-based BIM information retrieval. IFCMemoryBench contains 143 multi-session tasks across 19 projects and 4,016 prior sessions, derived from incomplete-information questions in IFC-Bench v2. Each task seeds missing project context across earlier conversations and later asks a probe question that can be answered only by combining remembered context with live IFC queries. Our evaluation framework decomposes memory performance into ingestion, retrieval, and utilization, and measures both answer quality and memory quality with expert-validated LLM judges.
We evaluate representative vector-, graph-, and file-based memory systems. The strongest system achieves only 32.4\% answer accuracy under a deployment-realistic ingestion scope, and remains below 60\% under oracle-filtered ingestion or a stronger probe agent. Analysis shows that current general-purpose memory systems often retrieve topically relevant context but store project knowledge as incomplete or fragmented facts. These results reveal a domain-transfer gap in agent memory and suggest that reliable professional agents require domain-aware memory representations linking conversations, project knowledge, and structured model entities.
\end{abstract}



\maketitle
\acmConference[KDD '26]{KDD 2026 Workshop on Evaluation and Trustworthiness of Agentic AI}{August 9-13, 2026}{Jeju, Republic of Korea}

\section{Introduction}

As LLM-based agents move from single-turn assistants toward sustained, tool-using systems, long-term memory is becoming a core deployment capability. Agents are increasingly expected to maintain project state, reuse information from prior interactions, and combine remembered context with live tool outputs. However, most existing evaluations still treat memory largely as conversational recall: whether an agent can recover facts, preferences, or past events from earlier dialogue. This setting is valuable, but it does not fully capture professional workflows, where remembered information must be used together with domain-specific tools, structured data, and evolving project context. In such settings, memory failures can lead not only to forgotten facts, but also to incomplete or misleading decisions.

We study this problem in Building Information Modelling (BIM), a professional engineering workflow that provides a concrete stress test for long-term memory in agentic AI. BIM models are structured digital representations of building projects, containing typed objects such as walls, beams, spaces, systems, materials, quantities, and relationships. Their dominant exchange format, the Industry Foundation Classes (IFC), is a large and semantically rich object-oriented schema. Querying IFC models requires domain expertise and technical skill, which has motivated recent work on LLM-based BIM information retrieval, where agents answer natural-language questions by inspecting IFC files through ReAct-style tool use~\cite{Gao2026, Guo2025, Hellin2026}.

Despite this progress, existing LLM-based BIM retrieval systems and benchmarks remain largely stateless. Benchmarks such as IFC-Bench~\cite{Hellin2026, Hellin2025} evaluate independent question-answer pairs, usually assuming that each answer can be derived from the IFC model alone. Real BIM workflows are different. Users return to the same project over weeks or months, refine requirements, confirm specifications, record client decisions, and correct earlier assumptions. Much of this information never enters the IFC model, yet later questions may depend on it. For example, a query may require IFC quantities together with an emission factor, cost assumption, room-schedule correction, or engineering convention mentioned in a previous session. A reliable BIM retrieval agent must therefore combine live IFC queries with long-term project memory.

This need reflects a broader gap in agent-memory evaluation. Existing memory systems use vector stores, knowledge graphs, or persistent files to retain information across sessions~\cite{Zhong2024, Chhikara2025, Rasmussen2025, Anthropic2025, LangChain2025}, and benchmarks such as LongMemEval~\cite{WuD2024}, MemoryAgentBench~\cite{Hu2026}, and MemoryArena~\cite{He2026} evaluate important memory capabilities in open-domain or persona-grounded conversations. However, they do not directly test whether memory remains reliable in a tool-grounded professional workflow, where the agent must combine remembered context with structured, domain-specific data. In such settings, retrieving topically relevant memories is not enough: the memory must preserve exact project facts, units, entity references, assumptions, and their relationship to the structured model.

To address this gap, we introduce IFCMemoryBench, a benchmark for evaluating long-term memory in LLM-based BIM information retrieval. IFCMemoryBench converts incomplete-information questions from IFC-Bench v2 into multi-session memory tasks. Each task seeds missing project context across earlier chat sessions and later asks a probe question that can be answered only by combining remembered context with live IFC queries. The benchmark contains 143 memory-dependent tasks across 19 projects and 23 IFC models, with 4,016 prior chat sessions in total.

Our evaluation decomposes memory into ingestion, retrieval, and utilization. Ingestion measures whether prior sessions are written into memory in a useful form; retrieval measures whether the agent can access the needed remembered context; and utilization measures whether the agent can combine that context with IFC query results to produce the final answer. We evaluate both answer quality and memory quality using expert-validated LLM judges.

We evaluate representative vector-, graph-, and file-based memory systems. The results reveal a substantial domain-transfer gap. Under a deployment-realistic ingestion scope that does not assume prior knowledge of which messages contain durable facts, the strongest system achieves only 32.4\% answer accuracy. Even with oracle-filtered ingestion or a stronger probe agent, accuracy remains below 60\%. Current systems often retrieve context that is topically relevant but incomplete, fragmented, or weakly connected to IFC entities and project assumptions. These findings suggest that reliable memory for professional agents requires domain-aware representations linking conversations, project knowledge, and structured model entities. This paper makes three contributions:

(1) We propose a methodology for converting stateless BIM information retrieval questions into multi-session, memory-dependent agent tasks.

(2) We introduce IFCMemoryBench, a benchmark of 143 long-term memory tasks across 19 projects and 4,016 prior sessions, together with an evaluation framework that measures both answer quality and memory quality.

(3) We provide a systematic study of representative vector-, graph-, and file-based memory systems for IFC-grounded retrieval, exposing a domain-transfer gap in current general-purpose memory systems.

\section{Background and Related Work}

\subsection{Memory in LLM-based agents}

The LLM-agent literature borrows memory
terms from cognitive science as engineering analogies for how long
information persists and how it is reused~\cite{Sumers2023, WuY2025}. Based on this taxonomy, we distinguish working,
short-term and long-term memory.
\textit{Working memory} denotes the task context
actively maintained while the agent handles a current user
request, such as the instruction, intermediate state and recent
tool outputs. \textit{Short-term memory} is the conversational history within one session: it
keeps the agent coherent across a few turns, typically by
appending past messages to the prompt or by summarising them.
\textit{Long-term memory} persists across sessions and is itself
usually subdivided into \textit{semantic} memory
(decontextualised facts, constraints, preferences), \textit{episodic} memory (specific past interactions and
events), and \textit{procedural} memory
(skills and routines).

Our benchmark uses this distinction only as interpretive
background. Operationally, each task tests whether the agent can
answer a BIM question by combining information read from the IFC
model with project context seeded in prior chat sessions. That context
may contain semantic-like project
facts from schedules or specifications (e.g.\ window glazing or
room finish requirements) as well as episodic-like conversational
updates (e.g.\ a room-schedule correction made in an earlier
session). Working and short-term memory concern context management
within the current task or session, whereas this study focuses on
long-term memory that persists across sessions; procedural skills
are left to future work.

\subsection{LLM-based BIM information retrieval}

Early LLM-based BIM information retrieval systems treated the task
as natural-language-to-query translation over IFC-derived data, as
in BIM-GPT~\cite{Zheng2023}. More recent work instead uses
ReAct-style agents that inspect IFC files through iterative tool
calls, observing each result before choosing the next query or
traversal. Gao et al.~\cite{Gao2026} introduce a schema-guided multi-agent
framework that factors the task into specialised agents reasoning
over the IFC schema and a shared tool layer. Guo et al.~\cite{Guo2025}
propose a multi-agent alignment framework that maps user queries
to BIM domain-specific language and library-code functions. Hellin
et al.~\cite{Hellin2026} generalise the trend to \textit{adaptive
exploration}, where the agent writes and executes Python code
against the IFC, and release IFC-Bench~v2.

A useful distinction across these systems is between
\textit{closed-world} and \textit{open-world} settings. Most
existing LLM-based BIM information retrieval work is
closed-world: the question is fully answerable from the IFC
itself, which cover direct property lookup and quantity calculations, etc.
Open-world questions require external context not contained in the IFC, such as a project
schedule, a structural specification, a designer's intent. Although such questions cannot be
answered from the IFC alone, they are central to practical AEC
workflows, where BIM queries often support design coordination,
compliance checking, and client-driven decisions. In an
agent-assisted workflow~\cite{Du2024Copilot, Deng2025BIMgent, Du2026Text2BIM}, this external context may be supplied by
the user during previous conversations with the agent. This suggests a
role for long-term agentic memory in retaining
information that is not present in the IFC itself. Our benchmark
targets this open-world regime.

\subsection{Agent memory systems and benchmarks}

Long-term memory systems persist selected information beyond the
context window so it remains available in later sessions. Practical implementations fall into three storage-oriented
designs. \textit{Vector-based} systems store extracted facts,
summaries, or interaction fragments as embeddings and retrieve them by
semantic similarity, as in MemoryBank~\cite{Zhong2024} and Mem0~\cite{Chhikara2025}. \textit{Graph-based} systems represent memory
as a knowledge graph that grows incrementally with the sessions and
supports hybrid retrieval~\cite{Rasmussen2025}. \textit{File-based}
systems keep memory as a plain text file that the agent reads for
context and edits to add, update, or delete information, as in the
Markdown \texttt{AGENT.md} files of Claude Code~\cite{Anthropic2025} and
the filesystem-backed memory in DeepAgents~\cite{LangChain2025}.

Long-term memory has been studied with a parallel line of
benchmarks. LongMemEval~\cite{WuD2024} evaluates chat
assistants on long-term interactive memory; MemoryAgentBench~\cite{Hu2026} examines memory along four capabilities
including accurate retrieval and conflict resolution; MemoryArena~\cite{He2026} focuses on interdependent multi-session
agentic tasks. However, these general-purpose memory benchmarks were not designed for
our domain.

\subsection{Research gaps}

Three gaps motivate this work. First, BIM retrieval benchmarks such
as IFC-Bench treat tasks as independent, stateless queries and do not
exercise long-term memory. Second, general memory benchmarks test
conversational or textual recall, not the use of remembered project
context alongside live tool-based queries over structured IFC data.
Third, it remains unclear how different long-term memory architectures
perform on IFC-grounded retrieval. We address these gaps with an
experimental framework for evaluating long-term agentic memory in BIM
information retrieval.

\section{Methodology}

\subsection{Benchmark design}
\label{sec:benchmark-design}

Each benchmark task asks an agent to answer a memory-dependent project
question $Q_t$ at probe time $t$. Before the probe, the agent has
observed the prior chat sessions $S_1,\dots,S_{t-1}$, from which a
memory system $f$ constructs a memory store. At probe time, the agent may also query the IFC model through query tool $T_{\mathrm{IFC}}$.
The produced answer $A_t$ is evaluated against a
gold answer $\hat{A}_t$ and the project information seeded in the prior
sessions. The benchmark is agnostic to how memory is represented or
accessed, requiring only that remembered project context can be used at
probe time. We model this process through three memory capabilities:
\begin{align}
  M_i    &= f_{\mathrm{ingest}}(M_{i-1},\, S_i),\quad i=1,\dots,t-1, \label{eq:ingest}\\
  R_t    &= \bigcup_{j=1}^{k} f_{\mathrm{retrieve}}(q_j,\, M_{t-1}) \;\subseteq\; M_{t-1}, \label{eq:retrieve}\\
  A_t    &= \mathrm{LLM}(Q_t,\, R_t,\, T_{\mathrm{IFC}}). \label{eq:utilize}
\end{align}
where $M_{t-1}$ is the memory store accumulated from the prior
sessions, and $R_t$ is the part of that store retrieved for the probe.
(1)~\textbf{\textit{Ingestion}} (Eq.~\ref{eq:ingest}) integrates
the prior sessions into memory \emph{sequentially}, merging each
session $S_i$ into the current store $M_{i-1}$ to preserve useful information
while filtering distractor questions, routine tool
traces, and conversational noise. This write step is \emph{online} and
\emph{stateful}: each session $S_i$ is compared with the
accumulated store $M_{i-1}$, and the system decides what to keep,
summarise, overwrite, or discard. This is what separates long-term
memory from retrieval-augmented generation (RAG): RAG indexing is
lossless and append-only, storing content without changing previous
entries and leaving selection to retrieval at query time. In contrast,
memory makes selective decisions during writing and integrates new
content into a store that evolves across sessions.
(2)~\textbf{\textit{Retrieval}}
(Eq.~\ref{eq:retrieve}) is the LLM agent dynamically querying a memory
retrieval function, and we distinguish two sides. \emph{System-side}
retrieval is the query function $f_{\mathrm{retrieve}}$ provided by the
memory system: given a query $q_j$, it returns a fixed subset of the
store $M_{t-1}$. \emph{Agent-side} retrieval is how the probe LLM agent
uses this function: the agent chooses how many queries $k$ to issue and
writes each $q_j$ itself, so $R_t$ is the union over the $k$ results.
Which part of $M_{t-1}$ this covers therefore depends on what the agent
decides to query. (3)~\textbf{\textit{Utilization}}
(Eq.~\ref{eq:utilize}) combines the retrieved memory $R_t$, the probe
question $Q_t$, and IFC results from $T_{\mathrm{IFC}}$ to produce $A_t$,
while keeping remembered context distinct from information read directly
from the IFC model. This separation lets us later diagnose failures according to whether
they arise in memory writing, memory retrieval, or answer-time use.

\subsection{Dataset construction}
\label{sec:dataset}

\begin{figure*}[t]
  \centering
  \includegraphics[width=\textwidth]{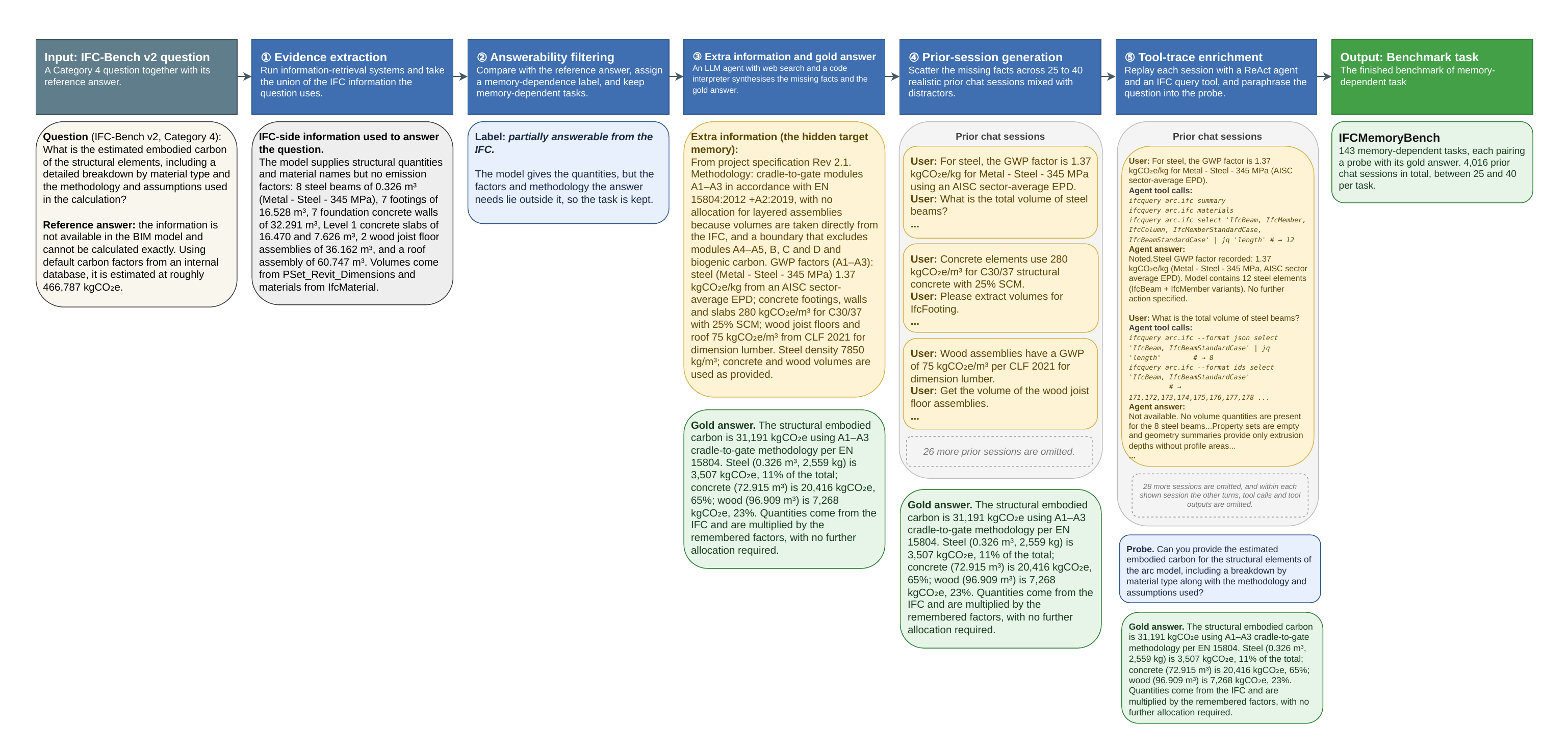}
  \caption{IFCMemoryBench construction pipeline. An IFC-Bench~v2
  Category~4 question is turned into a memory-dependent task, with one
  embodied-carbon example threaded through all stages.}
  \label{fig:pipeline}
\end{figure*}

The dataset is built on IFC-Bench~v2~\cite{Hellin2026}, which
groups questions by increasing information demand. Categories~1--3 are
closed-world tasks answerable from the IFC model itself, whereas
Category~4 covers incomplete-information cases in which the required
facts are absent, ambiguous, or must be estimated from sources outside
the model. We treat these Category~4 questions as seeds for
multi-session memory tasks: the missing external information is supplied
through earlier chat sessions, and the original question is asked only
afterwards as the probe. The remaining steps, summarised in
Figure~\ref{fig:pipeline}, describe how each task is built.

\textbf{Extracting the IFC-side information.} We run three
different high-performing information-retrieval systems on all
Category~4 questions from IFC-Bench~v2. Each run produces, for every question, a trace of the IFC
information retrieved while answering it. An LLM then
cross-analyses the three traces for each question to derive the
union of IFC information the question requires: which entities,
property sets, quantities, and spatial containments mattered. The
result is a precise account of what the IFC supplies.

\textbf{Filtering by answerability.} A second LLM compares the
extracted IFC information against the IFC-Bench~v2 ground-truth
answer and assigns one of three memory-dependent labels:
\texttt{not\_in\_ifc} (the IFC contains essentially nothing
relevant to the question), \texttt{needs\_external\_info} (the IFC supplies
context, but the answer hinges on information beyond the model),
and \texttt{partially\_\allowbreak answerable\_\allowbreak from\_\allowbreak ifc} (the IFC yields
part of the answer but leaves gaps or ambiguities). All three
labels signal a dependence on remembered context and are kept;
any question that proves fully answerable from the IFC alone is
discarded, since it would not exercise memory.

\textbf{Synthesising missing context and the gold answer.} For
each kept task, an LLM agent equipped with web search and a code
interpreter is given the assigned label, the extracted IFC
information, and the original IFC-Bench~v2 question, and produces
two outputs. The first is the \textit{extra information} a perfect
answer needs but the IFC lacks: the agent retrieves external general
facts such as standards, EPD/GWP emission factors and unit costs through
web search, and uses the code interpreter for aggregation or unit
conversion. Missing project-specific documents that cannot be looked up,
such as schedules or specifications, are instead synthesised as
plausible project context consistent with the IFC information. The
second output is the corresponding \textit{gold answer}, which combines
the extra information with the IFC context and serves as the
ground-truth answer for the probe. Figure~\ref{fig:pipeline}
illustrates this step with the embodied-carbon task. The
\textit{extra information} becomes the hidden target memory content.

\textbf{Generating realistic prior chat sessions.} A dataset-generation agent then synthesises between $25$ and $40$ prior chat
sessions per task. Each session contains a small set of user
messages that combine realistic BIM information retrieval queries
(about levels, walls, materials and schedules), natural distractors,
and natural reintroductions of the project knowledge derived from the
\textit{extra information}. The seeded information appears as
specifications mentioned in passing, user corrections to the agent, or
background context for unrelated requests, so that it is distributed
across sessions, wordings and surrounding dialogue. To convert these
messages into realistic human-agent conversation, we replay each synthetic session
against the same IFC model using a ReAct agent equipped with an IFC
query CLI tool. The agent answers turn by turn. At each turn, it can
access only the current user message, previous dialogue within the same
chat session, the IFC model through the CLI tool, and its own tool outputs.
This restriction prevents facts from other sessions, the hidden
\textit{extra information}, or the gold answer from leaking into the
generated dialogue. We store the complete interaction, including user
messages, assistant messages, tool calls and tool outputs, as the
session's \texttt{turns}. Finally, the
original Category~4 question is paraphrased into a final
\textit{probe question} that does not reveal the gold answer but
explicitly invites the agent to consult both memory and the IFC.

Every step of the dataset construction pipeline is validated by human review, which
discards any case where the gold answer is incorrect, where the
seeded information conflicts with the IFC, or where the probe is
ambiguous. The resulting benchmark contains 143 tasks. Each task
pairs a probe question with its gold answer and a set of prior
chat sessions where user messages scatter the project knowledge
needed to answer it, for a total of 4\,016 prior sessions across
the benchmark.

\subsection{Evaluation metrics}
\label{sec:eval-metrics}

Each system is evaluated along two independent axes:
\textit{answer quality} and \textit{memory quality}. Each task is
evaluated with an LLM-as-judge protocol using the probe, gold
answer, system response, the project information seeded in prior
sessions, and the memory retrieved during the run.

\textbf{Answer judge.} The answer judge assesses the final
response along three binary dimensions: (1)~\texttt{correctness},
whether its facts and conclusions match the gold answer, allowing
for paraphrase and small rounding differences; (2)~\texttt{completeness},
whether it provides the central result and any qualifiers needed
to avoid misleading the user; and (3)~\texttt{relevance}, whether
it addresses the probe rather than a neighbouring question,
project, or element. A task is counted as answer-correct only
when all three dimensions are true, and we report \emph{answer
accuracy} as the percentage of answer-correct tasks. This design follows existing evaluation practice in BIM
question answering and agent-memory benchmarks~\cite{Hellin2026, Hu2026}, and reduces sensitivity to verbosity
differences across memory systems.

\textbf{Memory judge.} The memory judge assesses how well the
project information seeded in prior sessions was preserved in
memory, retrieved at probe time, and used in the final response. It
scores three binary dimensions: (1)~\texttt{retrieval\_relevant},
whether the retrieved memory is about the right probe, project and
elements; (2)~\texttt{retrieval\_covers\_key\_facts}, whether the
retrieved memory preserves enough of the seeded project
information to supply the hidden facts needed for the answer; and
(3)~\texttt{answer\_uses\_memory}, whether the final answer uses
the retrieved memory rather than fabricating or relying only on
IFC information. A task is counted as memory-correct only when
all three dimensions are true, and we report \emph{memory
accuracy} as the percentage of memory-correct tasks.

\textbf{Human validation of the judges.} To confirm that both
judges track expert judgement, one domain expert independently
re-labelled a random sample of 40 tasks drawn across the system runs
of the main controlled comparison. On
this single sample the expert assigned, for each task, both an
answer-correctness and a memory-correctness label under the rubrics
of Sec.~\ref{sec:eval-metrics}. For answer-correctness the expert and
the judge agreed on 38 of the 40 tasks ($95\%$ raw agreement, Cohen's
$\kappa = 0.90$); for memory-correctness they agreed on all 40 tasks. Both values lie in
the almost-perfect band of the Landis--Koch scale. The two
answer-judge disagreements were boundary cases under the correctness
rubric rather than substantive errors: in one, the system answer
differed from the gold value only by a rounding/unit-precision margin
that the expert accepted but the judge did not; in the other, a
qualifier the expert considered necessary for completeness was
present only implicitly in the response. Such boundary cases are
specific to graded answer correctness; the memory dimensions instead
turn on whether the retrieved context is on-topic, covers the seeded
facts, and is actually used, which the expert and the judge read off
consistently, leaving no disagreements on this sample. Neither
answer-judge disagreement altered the relative ranking of the
affected systems, so we use the judge labels for all results reported
below.

\subsection{Studied memory systems}
\label{sec:systems}

All tested systems use the same ReAct-style BIM retrieval agent: the
agent receives the probe, a fixed prompt policy, and read-only
access to IFCQuery, a CLI for querying IFC
model information~\cite{IfcOpenShell2025}. We
vary only the attached memory layer: how prior sessions are
ingested, how relevant context is retrieved, and how that context
is exposed to the agent. A no-memory condition discards prior
sessions and serves as the lower bound/baseline.

\textbf{Vector memory.} The vector-memory condition instantiates
the Mem0 pattern~\cite{Chhikara2025}. During ingestion, prior
conversation messages are sequentially passed through a memory-extraction
pipeline that distils them into standalone textual memory items.
These items are embedded and stored in an item-scoped vector
database collection, together with metadata that separates one
benchmark task from another. During retrieval, the agent's query is
embedded and used for top-$k$ semantic similarity search over that
vector collection. The only memory surface exposed to the probe
agent is a \texttt{search\_project\_memory(query, limit)} tool that returns
the ranked memories as text.

\textbf{Temporal graph memory.} The graph-memory condition
instantiates the Graphiti temporal-knowledge-graph pattern~\cite{Rasmussen2025}. During ingestion, each prior session is
stored as a time-stamped episode, which acts as the source record
for facts extracted from that conversation. Graphiti then creates
nodes for important entities mentioned in the session and edges representing factual relationships between them, with temporal validity
fields where available. Retrieval uses hybrid search that combines
semantic and keyword matching over the graph, then surfaces
structured results such as relation facts, entity summaries and
validity timestamps. To keep the agent interface comparable with the vector
condition, these results are again exposed through
\texttt{search\_project\_memory(query, limit)}, but the returned
text is generated from graph records rather than from a flat vector
index.

\textbf{File-based memory.} The file-memory condition
instantiates an agent-managed Markdown memory pattern using
DeepAgents~\cite{LangChain2025}. During the ingestion,
a dedicated memory-writing agent processes the prior sessions one
at a time, deciding after each session whether to write or update a
compact project memory file (\texttt{memory.md}). We test
two file-memory forms. The uncited form exposes only the Markdown
summary. The cited form writes the same kind of Markdown summary but
adds stable message citations such as \texttt{[S003-M02]}, which
point into a companion \texttt{prior\_sessions.json} archive of the
original messages for optional \texttt{grep}/\texttt{read\_file}
verification. At probe time, \texttt{memory.md} is injected directly
into the probe agent's system prompt as read-only project context,
rather than retrieved through a search tool.

\section{Experiments and Results}
\label{sec:results}

\subsection{Benchmark statistics}

Table~\ref{tab:stats} summarises the final benchmark. The 143
tasks cover 19 projects and 23 IFC models spanning
architectural, structural, MEP, plumbing, ventilation and city
disciplines. The benchmark contains in total 9\,488 user messages,
45\,983 assistant messages and 39\,532 tool messages.
All project information seeded into each task are synthetic or derived from publicly available
sources. The IFC models are drawn from the public
IFC-Bench~\cite{Hellin2026, Hellin2025}.

\begin{table}[t]
  \centering
  \small
  \caption{IFCMemoryBench dataset statistics.}
  \label{tab:stats}
  \begin{tabular}{@{}lr@{}}
    \toprule
    Property & Value \\
    \midrule
    Tasks                            & 143 \\
    Projects                         & 19 \\
    IFC models                       & 23 \\
    Prior sessions                   & 4\,016 \\
    Prior sessions / task (avg.)     & 28 \\
    User messages                    & 9\,488 \\
    Assistant messages               & 45\,983 \\
    Tool messages                    & 39\,532 \\
    Total messages                   & 95\,003 \\
    Messages / session (avg.)        & 23.7 \\
    \bottomrule
  \end{tabular}
\end{table}

\subsection{Experimental setup}

All main experiments use the same ReAct probe agent implemented with
the DeepAgents framework. The agent is equipped with a bash
execution tool for running \texttt{ifcquery}; unrelated DeepAgents
tools are removed so that IFC access is mediated through this
interface. To keep the comparison fair, Grok-4.3 is used for the
probe agent, the LLM judges, and the LLM-based components inside
the tested memory systems, with temperature 0. We instantiate each memory system with its standard default settings except where noted.
Mem0 stores its memory items in a Qdrant vector database with the
default embedding model. Graphiti is backed by Neo4j; we keep its
default embedder, graph schema, and hybrid retrieval, which fuses
lexical (BM25) and dense semantic candidates by reciprocal-rank fusion
(RRF). The file memory is built on the DeepAgents
\texttt{MemoryMiddleware}, which reloads \texttt{memory.md} from disk
at each step.

\subsection{Memory system comparison}
\label{sec:memory-system-comparison}

The decomposition in Sec.~\ref{sec:benchmark-design} also sets the
control logic for the experiments. In this main comparison, we fix the
probe agent and vary the memory layer, so observed differences are
attributed to how each memory architecture stores and exposes project
context to the same agent. Section~\ref{appendix:llm-effect} uses the complementary control: it
fixes the stored memory and varies the probe LLM to examine how agent
capability affects agent-side retrieval and memory use.

This main comparison fixes the ingestion scope to user messages plus the
assistant's final answer, rather than the user-messages-only scope that
scores highest in Sec.~\ref{sec:ingestion-scope}. This reflects
deployment realism: a memory layer running over a live conversation
cannot know in advance which turns carry durable project facts, whereas
the user-only scope presupposes a perfect pre-filter and acts as a
near-oracle ingestion scope rather than an operating point a deployed
system could assume (not the full-context oracle of Sec.~\ref{sec:oracle},
which bypasses the memory layer entirely). Reporting the deployment-faithful
scope as the headline gives a conservative estimate; Sec.~\ref{sec:ingestion-scope}
then isolates how much of the gap is attributable to ingestion scope alone.

Table~\ref{tab:main} reports answer and memory-judge scores. We
analyse the results from three perspectives: how prior sessions are
ingested into memory, how useful memory is retrieved or surfaced at
probe time, and how the agent uses that memory in the final answer.
The no-memory lower bound answers no task
fully, confirming that the benchmark requires cross-session memory;
yet even the best system under this fixed-probe setting reaches only
32.4\% answer accuracy. The Wilson 95\% intervals around these
accuracies overlap for several systems, so we do not over-interpret
small gaps in the ranking; what the intervals make unambiguous is that
every system performs poorly in absolute terms, with all upper bounds
far below the level a reliable assistant would need. We therefore ask
where failures become visible across memory ingestion, retrieval, and
utilization in the final answer.

\begin{table*}[t]
  \centering
  \small
  \caption{Per-dimension results on IFCMemoryBench. Each value is the
  percentage of tasks for which the corresponding binary (yes/no)
  judge returns positive. The answer judge scores correctness
  (\textit{Corr.}), completeness (\textit{Cmp.}) and relevance
  (\textit{Rel.}); answer accuracy (\textbf{Ans-acc.}) requires all
  three to hold. The memory judge scores whether the retrieved context
  is relevant (\textit{R-rel.}), covers the key facts (\textit{R-cov.})
  and is actually used in the answer (\textit{A-mem.}); memory accuracy
  (\textbf{Mem-acc.}) requires all three to hold. Bracketed values under
  each rate are Wilson score 95\% confidence intervals for the underlying
  binomial proportion over the evaluated tasks~\cite{Wilson1927}.}
  \label{tab:main}
  \resizebox{\textwidth}{!}{%
  \begin{tabular}{@{}lcccc@{\hspace{2.2em}}cccc@{}}
    \toprule
    & \multicolumn{4}{c}{Answer judge (\%)} & \multicolumn{4}{c}{Memory judge (\%)} \\
    \cmidrule(lr){2-5}\cmidrule(lr){6-9}
    System & Corr.\ & Cmp.\ & Rel.\ & \textbf{Ans-acc.} & R-rel.\ & R-cov.\ & A-mem.\ & \textbf{Mem-acc.} \\
    \midrule
    No memory             & \acc{22.4}{16.3}{29.9} & \acc{3.5}{1.5}{7.9} & \acc{92.3}{86.8}{95.7} & \acc{0.0}{0.0}{2.6} & ---  & ---  & ---  & ---  \\
    Graphiti (graph)      & \acc{32.2}{25.1}{40.2} & \acc{22.4}{16.3}{29.9} & \acc{99.3}{96.1}{99.9} & \acc{21.0}{15.1}{28.4} & \acc{81.8}{74.7}{87.3} & \acc{27.3}{20.6}{35.1} & \acc{74.1}{66.4}{80.6} & \acc{25.9}{19.4}{33.6} \\
    Markdown uncited (file) & \acc{25.4}{18.9}{33.1} & \acc{23.2}{17.1}{30.8} & \acc{100.0}{97.4}{100.0} & \acc{17.6}{12.2}{24.7} & \acc{66.9}{58.8}{74.1} & \acc{24.6}{18.3}{32.3} & \acc{61.3}{53.1}{68.9} & \acc{24.6}{18.3}{32.3} \\
    Markdown cited (file)   & \acc{30.1}{23.2}{38.0} & \acc{30.1}{23.2}{38.0} & \acc{100.0}{97.4}{100.0} & \acc{24.5}{18.2}{32.1} & \acc{70.6}{62.7}{77.5} & \acc{43.4}{35.5}{51.5} & \acc{67.1}{59.1}{74.3} & \acc{38.5}{30.9}{46.6} \\
    Mem0 (vector)         & \acc{\textbf{44.4}}{36.4}{52.6} & \acc{\textbf{33.1}}{25.9}{41.2} & \acc{100.0}{97.4}{100.0} & \acc{\textbf{32.4}}{25.2}{40.5} & \acc{\textbf{90.1}}{84.1}{94.0} & \acc{\textbf{49.3}}{41.2}{57.4} & \acc{\textbf{82.4}}{75.3}{87.8} & \acc{\textbf{47.2}}{39.2}{55.4} \\
    \bottomrule
  \end{tabular}}
\end{table*}

The first comparison concerns what is written from prior sessions.
Under the fixed Grok probe agent, Mem0 performs best overall, but the
two Markdown variants isolate the effect of written memory quality
because they share the same storage and injection mechanism. The cited
variant has higher observed scores than the uncited one, suggesting that memory
should preserve traceable, self-contained project facts rather than
compressed session summaries.
Graphiti shows the same issue from another angle: relevant graph
records do not necessarily expose facts in a form the agent can reuse.

The main bottleneck is access coverage. Across mechanisms, the agent
usually receives context from the right project and topic, but that
context often omits key facts. This relevance--coverage gap is clearest
for Graphiti, where specification-level facts may be split across nodes
and edges instead of returned as coherent evidence. Mem0 and cited
Markdown reduce this problem, but even the best system covers only
about half of the required facts.

Finally, under this fixed agent, \texttt{A-mem.} indicates that useful
memory is often used when it is available. Since these scores exceed
coverage scores, the immediate bottleneck in the main comparison is
often missing facts in the accessed context rather than wholesale
failure to use memory. This also explains why answer relevance is high
while completeness remains much lower. Overall, the remaining failures
mainly reflect weaknesses in memory representation and access, though
Sec.~\ref{appendix:llm-effect} shows that agent-side search effort can also materially
change coverage and answer quality.

\subsection{Effect of ingestion scope}
\label{sec:ingestion-scope}

Reliable cross-session reuse depends not only on the memory architecture
but also on what is written to memory from each prior session. We
therefore hold storage, retrieval and probe fixed, and vary only the
ingestion scope: only user messages, user messages plus the
assistant's final answer, and full turns
including intermediate reasoning, tool calls and outputs. Across all
three systems, the best results come from ingesting only user messages
(Table~\ref{tab:ingest}). This finding is partly a consequence of the
benchmark design, and should therefore be read with that limitation in
mind. In our construction (Sec.~\ref{sec:dataset}), durable facts are seeded only in
user messages; assistant and tool turns do not introduce additional
durable facts. Wider scopes therefore add mostly distractors. As this
noise increases, Mem0 is the most stable, Graphiti degrades when
assistant and tool text is converted into extra graph nodes, and the
Markdown file memory is least robust.

The cost effect is also substantial: per task, ingestion grows from
0.9\,k tokens with user messages to 10.2\,k with final
answers and 167\,k with full turns (180x).
Although ingestion time usually follows this increase, successful
full-turn runs average only about 215\,s because the agent often writes
no memory file under heavy noise. Full turns thus pair the highest read
cost with ingestion collapse, making scope selection a central design
choice rather than neutral preprocessing.

\begin{table}[h!]
  \centering
  \small
  \caption{Effect of ingestion scope. Quality is answer accuracy and memory accuracy (as in
  Table~\ref{tab:main}); ingestion cost is the tokens read per task and the mean per-task ingestion time. Full turns was
  run only for the file memory due to budget constraints.}
  \label{tab:ingest}
  \resizebox{\linewidth}{!}{%
  \begin{tabular}{@{}llcc@{\hspace{1.4em}}cc@{}}
    \toprule
    & & \multicolumn{2}{c}{Quality (\%)} & \multicolumn{2}{c}{Ingestion cost} \\
    \cmidrule(lr){3-4}\cmidrule(lr){5-6}
    System & Scope & \textbf{Ans-acc} & \textbf{Mem-acc} & token/task & time/task \\
    \midrule
    \multirow{2}{*}{Mem0}
      & user-messages & \textbf{52.4} & \textbf{60.8} & 0.9\,k & 104\,s \\
      & user-messages + agent-final-answer & 32.4 & 47.2 & 10.2\,k & 405\,s \\
    \midrule
    \multirow{2}{*}{Graphiti}
      & user-messages & \textbf{42.7} & \textbf{45.5} & 0.9\,k & 42\,s \\
      & user-messages + agent-final-answer & 21.0 & 25.9 & 10.2\,k & 110\,s \\
    \midrule
    \multirow{3}{*}{\shortstack[l]{Markdown\\cited (file)}}
      & user-messages & \textbf{54.4} & \textbf{62.5} & 0.9\,k & 191\,s \\
      & user-messages + agent-final-answer & 24.5 & 38.5 & 10.2\,k & 374\,s \\
      & full-turns      & 16.5 & 27.8 & 167\,k & 215\,s \\
    \bottomrule
  \end{tabular}}
\end{table}

\subsection{Full-context oracle as a costed upper bound}
\label{sec:oracle}

We additionally evaluate full-context oracle conditions in which the probe agent
receives prior messages directly in its prompt, without memory extraction or retrieval.
These conditions are not intended as deployable memory baselines: they assume access
to the task-relevant prior sessions and, in the cleanest setting, to the message type
that contains the seeded durable facts. Instead, they serve as sanity-check upper bounds
on task answerability and make the cost model of long-context prompting explicit.

\begin{figure}[h!]
  \centering
  \includegraphics[width=\linewidth]{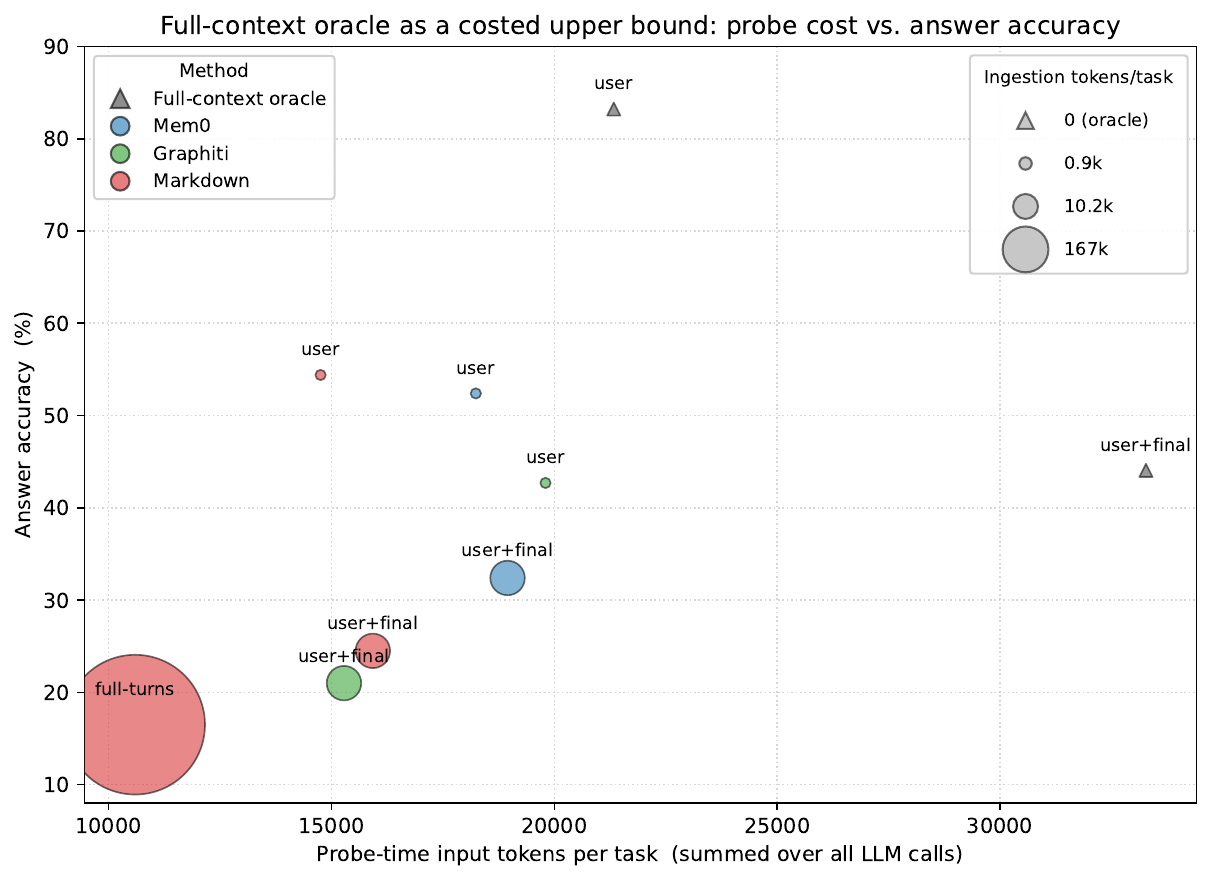}
  \caption{Cost--quality trade-off between full-context prompting and memory systems.
  The x-axis shows probe-time input tokens per task, the y-axis shows answer
  accuracy, and bubble area denotes one-time ingestion tokens per task.
  Full-context prompting gives a costed upper bound, while memory systems
  amortize ingestion across future queries but currently lose substantial
  quality during memory writing and access.}
  \label{fig:oracle}
\end{figure}

When all prior user messages are exposed losslessly, memory accuracy reaches 95.1\%
and answer accuracy reaches 83.2\%, showing that most benchmark tasks are solvable
when the relevant project facts are directly visible to the agent. However, as shown in
Figure~\ref{fig:oracle}, this accuracy is obtained by paying the history cost at probe time for every
query. In contrast, memory systems pay an up-front ingestion cost and expose a much
smaller context at probe time, allowing this write cost to be amortized across future
project queries.

The oracle also shows that more context is not automatically better. When we expand
the prompt from prior user messages to user messages plus assistant final answers,
memory accuracy drops from 95.1\% to 74.1\%, and answer accuracy drops from 83.2\%
to 44.1\%. This degradation indicates that naive long-context prompting is sensitive to
conversational noise: additional assistant-side text can obscure, dilute, or conflict with
durable project facts. The gap between the clean user-message oracle and the tested
memory systems therefore quantifies the loss introduced by current memory-writing,
representation, and access mechanisms, rather than an inherent ambiguity in the benchmark.

\subsection{Effect of domain customisation}

The experiments so far use each memory system's default, general-purpose
ingestion policy. We ask whether a light domain customisation, telling
the memory layer what an AEC/IFC memory should contain, narrows the
coverage gap without changing the architecture or retrieval. Holding the
agent, judges, model, retrieval and ingestion scope (user\,$+$\,final
answer) fixed, we apply one minimal customisation per system: for Mem0,
a domain instruction in its fact-extraction prompt that keeps exact
quantities, units and element or material names as self-contained facts;
for the DeepAgents file memory, a domain template that separates IFC-read facts
from user-supplied facts and records each external value atomically
under a fixed schema. We customise only these two systems, as Graphiti's
equivalent adaptation requires redesigning its entity and edge ontology
rather than a prompt, which we leave to future work.

\begin{figure}[t]
  \centering
  \includegraphics[width=\linewidth]{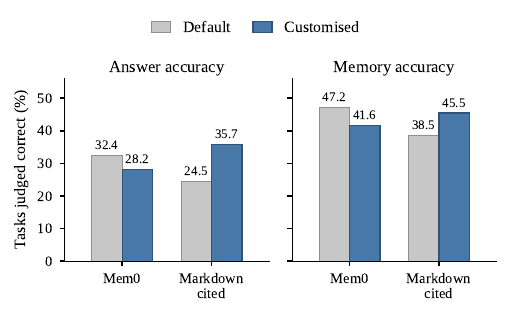}
  \caption{Effect of domain customisation.}
  \label{fig:domain}
\end{figure}

The two systems respond in opposite directions (Figure~\ref{fig:domain}).
In our experiments, the domain template was the more effective of the two
memory-layer changes for the agent-managed file memory: it raised both
answer and memory accuracy, bringing the file memory roughly level with
the default Mem0 store, apparently by making the retrieved context more
relevant and more complete. The same
instruction does not help Mem0 and slightly lowers both scores, since its
pipeline already distils standalone facts and the extra rule mainly
fragments each specification, reducing per-item coverage. The value of
customisation thus depends on the default ingestion quality: a weak,
generic writing policy benefits clearly, whereas an already strong one
gains little and can regress.

\subsection{Effect of LLMs used in the probe agent}
\label{appendix:llm-effect}

The main study fixes the probe agent, the judges and the
memory-system LLMs to Grok-4.3 for a fair comparison across architectures
(Sec.~\ref{sec:results}); here we change only the probe agent's LLM to
Gemini-3.5-Flash and hold everything else fixed, reusing the previously
ingested memory verbatim. Any change in scores therefore reflects the effect of
the probe LLM on agent-side retrieval and memory \emph{utilization}, not on how
the memory was written.

Each of the 143 tasks is answered by both agents from the same fixed memory, so
the two probe LLMs are compared on identical items as 143 \emph{paired}
correct/incorrect outcomes. We test a single quantity: the accuracy gain
$\Delta = p_{\mathrm{Gemini}} - p_{\mathrm{Grok}}$ between the two LLMs on these
items, whose posterior we estimate with a Bayesian paired test of proportions~\cite{Vosseler2026} (exact P\'olya--Gamma sampler, so no large-sample normal
approximation is needed at $n{=}143$). From the posterior we report the mean
$\Delta$, its 95\% credible interval (CrI) and $P(\mathrm{Gemini}>\mathrm{Grok})$.
We call a gain \emph{decisive} when its 95\% CrI excludes $0$ (its plausible
range lies entirely on one side of $0$, so the sign of the gain is settled),
and \emph{undecided} when the CrI still includes $0$ (the data do not yet
resolve whether the gain is real). We use a \emph{paired} test because
both agents see the same tasks, so pairing cancels task-difficulty noise, and a
Bayesian one because the posterior gives a directly interpretable probability
and effect size rather than a single $p$-value.

\begin{figure}[h!]
  \centering
  \includegraphics[width=\linewidth]{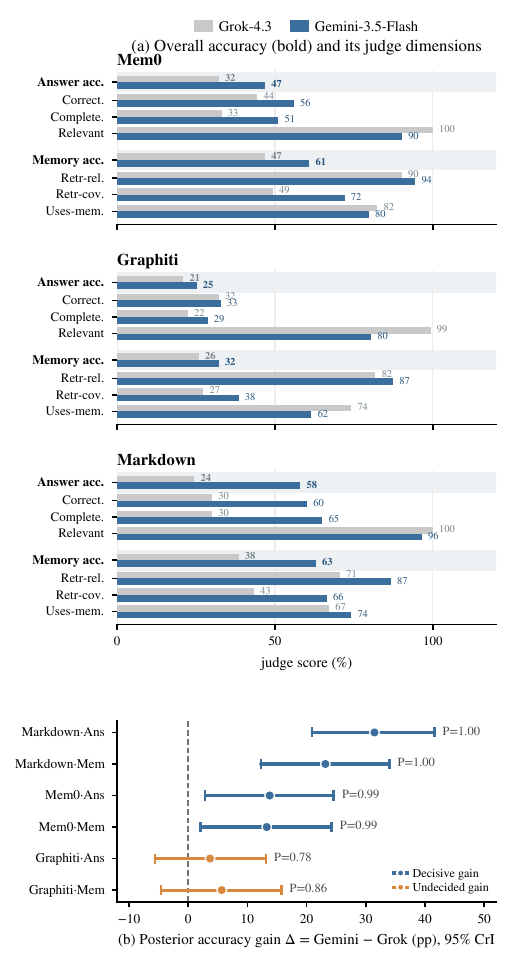}
  \caption{Effect of the probe-agent LLM, with memory held fixed.
  (a)~Overall answer and memory accuracy and the judge
  dimensions that compose them.
  (b)~Posterior accuracy gain $\Delta$ (Gemini\,$-$\,Grok), one row per system
  and metric. The dot is the posterior mean, the
  bar its 95\% credible interval (CrI), and the dashed line marks $\Delta{=}0$
  (no difference).}
  \label{fig:llm}
\end{figure}

Gemini improves both answer and memory accuracy
on all three systems (Figure~\ref{fig:llm} (b)): decisively for the file memory
and Mem0, and positively but undecidedly
for Graphiti, whose 95\% CrI still includes $0$. The gain even reorders the systems:
the file memory leads under Gemini, whereas the Mem0 store leads under Grok
(Table~\ref{tab:main}), so the ranking of memory architectures depends on the
agent, not on the memory layer alone.

The cause is search effort: Gemini issues several times more memory queries
than Grok (e.g.\ 8.1 vs.\ 1.7 per task for Mem0) and surfaces more context, at
a several-fold latency cost. As Figure~\ref{fig:llm} (a) shows, this mainly raises
the completeness and retrieval-coverage dimensions, while relevance is the only
dimension that falls; since each overall accuracy (the bold rows) is the
conjunction of its three dimensions, it rises only as far as its weakest
dimension allows. The file memory gains most, as its content is injected in
full and its citations can be verified on demand, so all of its dimensions
improve together; the graph gains least, its facts staying fragmented so that
retrieval-coverage remains low even for Gemini. A more capable agent thus
narrows but does not close the coverage gap, confirming that utilization, not
only ingestion, is a primary lever, and that the probe LLM should be reported
as a first-class experimental factor.

\section{Discussion}
\label{sec:discussion}

IFCMemoryBench is a synthetic, human-validated benchmark
whose design rests on a few assumptions.
First, task-relevant prior knowledge is seeded only through user chat
messages (Sec.~\ref{sec:dataset}), while real projects may also rely on
uploaded files, drawings or other external artefacts; extending the
benchmark to these sources is an important direction for future work.
Second, as Sec.~\ref{sec:ingestion-scope} notes, user-only ingestion
scoring best is partly an artefact of seeding durable facts only in user
turns. This limits ecological validity: in real projects,
assistant replies and tool outputs can themselves carry memorable
content, such as computed quantities, intermediate results, or
established query routines, that a deployed memory system would need to
retain. A future version of the benchmark will therefore seed durable
facts into assistant and tool turns as well, turning ingestion-scope
selection into a genuine selective-memory problem rather than a
near-trivial ``keep only user messages'' rule, and enabling evaluation
of procedural memory such as reusable IFC query routines.
Third, a single model family (Grok-4.3) simultaneously serves as the
probe agent, the LLM components inside the memory systems, and the
answer and memory judges, which in principle exposes the study to
self-preference bias, the tendency of an LLM judge to favour outputs
from its own family. In the main comparison this exposure is uniform
across systems, since every answer is produced by the same agent, so it
does not distort the relative ranking of memory layers; it is most
relevant to the probe-LLM comparison (Sec.~\ref{appendix:llm-effect}),
where a Grok judge scores both Grok and Gemini agents. Two design
choices bound the risk: every judgement is anchored to a human-authored
gold answer and a fixed binary rubric rather than to open-ended
preference, and the judges were validated against a blind expert
relabelling with almost-perfect agreement (Sec.~\ref{sec:eval-metrics}). Cross-model judging
remains a valuable robustness check for future work. We thus view
IFCMemoryBench as, to our knowledge, the first domain-specific
long-term-memory benchmark for BIM information retrieval, and as an
extensible basis for broader sources and memory types.

A further question is whether a dedicated memory layer is needed at all,
since long-context models ingest hundreds of thousands of tokens and the
prior sessions per task here are small enough to concatenate directly into
the prompt. Our full-context oracle makes this concrete: exposing all prior
user messages losslessly reaches 83.2\% answer accuracy
(Sec.~\ref{sec:oracle}, Fig.~\ref{fig:oracle}). We read this as a property
of the present task instances, not evidence against memory systems. First,
the benchmark scales beyond any fixed context window: adding prior sessions
and synthetic distractors degrades the relevant-to-irrelevant ratio
arbitrarily while the answer still turns on a few seeded facts, and our
full-turn ingestion result already shows indiscriminate ingestion
collapsing at the upper end (about 167k tokens per task,
Sec.~\ref{sec:ingestion-scope}). Second, production deployments accumulate
effectively unbounded history across months and many concurrent users,
which cannot be reloaded in full on every query for cost and latency
reasons: the oracle pays this probe-time cost on every query, whereas
memory systems amortize it through up-front ingestion. Selective storage
and retrieval are therefore required, and the bottleneck the benchmark
exposes, surfacing the right project facts, is exactly what a long-context
dump does not resolve. We thus treat the full-context oracle as a costed
upper bound and sanity check, not a substitute for the memory capability
under study.

Within this scoped setting, the experiments surface a set of interacting
challenges that span memory ingestion, access, and utilization. We frame
these as potential limitations and areas for improvement that the
benchmark makes visible, rather than as inherent failures of the systems
studied. The main experiments suggest that ingestion and representation
are the primary upstream limitation: project knowledge may be written
incompletely, inverted into an ``absent from the model'' finding, or
fragmented into isolated facts that are harder to use for list,
comparison or aggregation probes. These writing limitations in turn
appear to constrain retrieval coverage: agents often access topically
relevant memory, but not always the complete project facts needed for
the answer. Section~\ref{appendix:llm-effect} adds that, with stored memory held fixed,
the probe LLM still materially affects agent-side retrieval and memory
use through its search effort and evidence verification. Taken together,
these observations suggest that the challenge is less about retrieving
more text and more about preserving project knowledge in a structured,
traceable form and pairing it with agents capable of using that memory
alongside the live IFC.

We regard surfacing such limitations as a core value of the study rather
than a negative result. Because IFCMemoryBench is a controlled research
instrument applied to research prototypes, it can isolate where current
general-purpose memory systems lose project knowledge in a low-risk
setting, before such behaviour would matter in practice. Making these
areas for improvement explicit and measurable is what allows the
community to target them, and we hope the benchmark can serve as a shared
diagnostic for tracking progress on domain-grounded memory.

\section{Conclusions}

This paper introduced IFCMemoryBench, a benchmark and
extensible methodology for synthesising multi-session
memory tasks from existing BIM question-answering data, and used it for an
exploratory study of representative general-purpose memory systems in
IFC-grounded BIM information retrieval, intended to highlight areas for
improvement of existing systems. Our
observations point to a domain-transfer opportunity: such systems can often
retrieve related context, but have more room to preserve project knowledge
as complete, grounded evidence usable together with live IFC queries,
suggesting that BIM memory may benefit from being organised around the
structure of project information rather than treated as a generic
conversation-memory layer. The probe-LLM comparison further shows these rankings are
conditioned on the agent, making the LLM itself an important experimental
factor. We offer domain-adapted memory designs that connect conversations,
project documents and IFC entities, so that remembered knowledge could
better support traceable, measurement-based reasoning, as an open research
direction.

\begin{acks}
The authors would like to thank Sylvain Hellin and Stefan Fuchs for their valuable
suggestions and inspiring discussions.
\end{acks}

\bibliographystyle{ACM-Reference-Format}
\bibliography{ifcmemorybench}

\appendix

\section{Probe-Agent System Prompt}
\label{appendix:system-prompt}

For completeness and reproducibility, we reproduce below the verbatim
system prompt used by the ReAct probe agent (the DeepAgents-based BIM
retrieval agent of Sec.~\ref{sec:systems}). The same prompt policy is
shared across all memory conditions; only the attached memory layer
varies.

\begin{lstlisting}[style=promptstyle]
You are a BIM expert. Answer questions about a building project using two
available evidence sources:

- `ifcquery`, via the `execute` tool, for facts that are actually present in
  the IFC model.
- Project memory for remembered project facts, requirements, schedules,
  specifications, assumptions, decisions, corrections, and negative
  confirmations.

## Source Priority
- Do NOT treat the IFC model as the only source. Some questions are
  intentionally answerable from prior project memory even when the IFC model
  lacks the relevant schedule, specification, decision, or qualification.
- Read the probe question and the Project Memory before deciding whether the
  answer needs IFC exploration, project memory, or both.
- Use `ifcquery` when answering claims about IFC-model entities, quantities,
  geometry, spatial containment, property sets, materials, or model metadata.
- If prior-session memory directly answers a non-IFC project/specification/
  schedule/decision question, you may answer from memory without forcing an
  irrelevant IFC lookup first.
...

## The ifcquery CLI
You have two tools:
- `execute(command, timeout=...)` - runs a shell command in the project
  directory (cwd is already set to the directory containing the IFC file).
- `read_file(path)` - reads a file from the project directory. Supports
  `offset` and `limit` parameters for paginated reading of large files.

**Allowed commands:** use `ifcquery` for IFC queries and `jq` for JSON
post-processing. For bounded shell pipelines you may also use these helpers
only: `cat`, `head`, `tail`, `wc`, `sort`, `uniq`, `xargs`, and `sh -c`.
...

### Available Subcommands (read-only; all default to `--format json`)
- `summary` - schema version, total entity count, project metadata,
  per-class counts. When IFC exploration is needed, start here unless a more
  targeted query is clearly sufficient.
- `tree` - full spatial hierarchy as nested JSON
  (Project -> Site -> Building -> Storey -> elements). Can be very large;
  prefer `select` first if you only need one class.
- `info <step_id>` - deep inspection of any entity by its step ID.
  Returns: `id`, `type`, `attributes`, and (when applicable), `property_sets`,
  `element_type` (type definition), `material`, `container` (spatial parent),
  `placement` (4x4 matrix), `geometry_summary` (representation type, solid
  profiles, extrusion depths). Keys that are empty are omitted.
  `step_id` is the positive integer `id` from any select/tree/relations result.
- `select '<selector>'` - IfcOpenShell selector syntax. Common uses:
  'IfcWall', 'IfcWall, IfcSlab', 'IfcBuildingStorey'.
  Result: list of {id, type, repr, name}, sorted by id.
- `relations <step_id>` - relationships for an element. Returns `id`,
  `type`, `name`, and (when applicable) `hierarchy` (parent, container,
  filled_void), `children` (openings), `type_relationship` (type_of),
  `material`, `elements` (all related entities as a flat list).
  Keys that are empty are omitted.
- `relations <step_id> --traverse up` - walks spatial hierarchy from
  element up to IfcProject (returns ordered list).
...

### Output Format Flag
`--format json|text|ids` - place BEFORE the subcommand (default: `json`).
Example: `ifcquery {ifc_filename} --format ids select 'IfcWall'`
- `json` - structured JSON, best for parsing.
- `text` - indented human-readable output.
- `ids` - comma-separated step IDs, useful for chaining commands.
...

## IFC Exploration Strategy
Use this strategy when the question requires IFC-model evidence:
1. **Orient:** Run `ifcquery {ifc_filename} summary` to get schema version,
   entity counts, and project metadata. This also reveals units.
2. **Discover elements:** Use `ifcquery {ifc_filename} select '<IfcClass>'`
   to find elements of a specific type. For example:
   `ifcquery {ifc_filename} select 'IfcWall, IfcWallStandardCase'`
3. **Inspect details:** Use `ifcquery {ifc_filename} info <step_id>` to get
   full attributes, property sets, element type, material, container,
   placement, and geometry summary for any element.
...

## Important ifcquery Tips
- **Always use `{ifc_filename}` verbatim** - your cwd is already the
  project directory.
- **Prefer `--format json`** (default). Use `--format ids` only when you
  plan to pass step IDs into another command. Place the flag BEFORE the
  subcommand: `ifcquery {ifc_filename} --format ids select 'IfcWall'`.
- **Filter big outputs with `jq`**:
  `ifcquery {ifc_filename} select 'IfcWall' | jq 'length'` gives a count.
  `jq` is available in PATH.
- ...

## IFC Domain Knowledge
**IFC class variants:** Some IFC schemas split classes into
variants. When searching for a class, query ALL variants:
- Walls: `IfcWall` AND `IfcWallStandardCase`
- Slabs: `IfcSlab` AND `IfcSlabStandardCase`
- Beams/Columns: `IfcBeam`/`IfcColumn` AND `IfcBeamStandardCase`/`IfcColumnStandardCase`
- Members: `IfcMember` AND `IfcMemberStandardCase`
Common architectural classes: IfcWall, IfcDoor, IfcWindow, IfcSlab, IfcColumn,
IfcBeam, IfcStair, IfcRailing, IfcCovering, IfcCurtainWall, IfcPlate, IfcMember,
IfcFurnishingElement, IfcBuildingElementProxy.
Common MEP classes: IfcFlowSegment (pipes, ducts), IfcFlowTerminal (fixtures,
outlets, sensors), IfcFlowFitting (junctions, bends, tees), IfcFlowController
(valves, dampers, switches), IfcEnergyConversionDevice (boilers, chillers),
IfcDistributionFlowElement, IfcDistributionElement.
**MEP element subtyping:** In IFC2x3, specific terminal types (sanitary,
waste, fire suppression) are often stored as `IfcFlowTerminal` instances.
Their specific category is in the type definition's class (e.g.
`IfcWasteTerminalType`, `IfcSanitaryTerminalType`). 
...

## Unit Handling
**Check units first for IFC measurement questions.** Run
`ifcquery {ifc_filename} summary` and examine the `schema` and project
metadata for unit information. If `summary` does not include explicit
unit data, use `ifcquery {ifc_filename} select 'IfcUnitAssignment'`
followed by `info` on the result to discover units.
...

## Verification
Before giving your final answer, sanity-check:
- Did you use the source(s) the question actually requires: memory, IFC, or both?
- Are the units correct for IFC measurements?
- If the IFC lacks data, did you try reasonable alternative class names, type
  variants, and relationship patterns before saying IFC lacks it?
- If prior-session memory contains the missing fact, did you use that memory
  instead of answering only "not available in the model"?
...

## Answer Rules
- Base IFC-derived claims on data returned by ifcquery. Do NOT guess or infer
  IFC values that are not explicitly present.
- Base remembered project facts on project memory. Do NOT invent facts that are
  not present in the available memory.
- When counting or aggregating IFC data, cover ALL storeys and spatial
  structures.
- If information is genuinely absent from the IFC model, state that clearly as
  an IFC limitation, but still use prior-session memory if it supplies the
  requested project fact.
...

## Answer Quality
A good answer has four properties: it is **accurate** (numbers and facts are
correct), **complete** (all relevant items included), **transparent** (source
type is clear), and **relevant** (addresses exactly what was asked).
Keep source attribution concise. For final answers, identify whether key facts
come from prior-session memory, IFC model evidence, or both. Do not dump long
command strings or intermediate query details unless they are necessary to
disambiguate the answer.

...

Be direct and concise. No follow-up offers.
\end{lstlisting}

\section{LLM-as-Judge Prompts}
\label{appendix:judge-prompts}

We reproduce below the verbatim instruction prompts for the two
LLM-as-judge evaluators of Sec.~\ref{sec:eval-metrics}. Both judges
additionally receive the probe, the gold answer, the system response,
the project information seeded in prior sessions, and (for the memory
judge) the memory retrieved during the run; each returns one binary
verdict per dimension under a fixed structured schema.

\subsection{Answer Judge}

\begin{lstlisting}[style=promptstyle]
You are an expert evaluator. Your task is to judge whether a system answer correctly answers the probe question
using the gold answer as the target answer. You may receive project and ifc_model
only to disambiguate the project/model scope; do not use them as evidence for
facts that are not present in the gold answer.

Evaluate these three answer-accuracy dimensions independently:

1. correctness: The system answer's stated facts and conclusions are compatible
   with the gold answer. Paraphrases are acceptable. Mark correctness false for
   wrong yes/no conclusions, wrong project/model/scope, contradictory source
   claims, materially wrong numeric values, incompatible units, wrong list items,
   or extra factual claims that materially change the answer.
   Do not mark correctness false merely because the answer is less detailed than
   the gold answer; missing details are a completeness issue unless the omission
   changes or negates the requested conclusion. Accept small numeric differences
   caused by rounding, formatting, or precision, including values within about 2%
   of the gold value or conventional rounded/unrounded equivalents such as
   100.2 kW vs 100 kW. Discrete counts should normally match unless the answer
   explicitly gives an approximation and the difference does not affect the
   conclusion.
   Additional factual details should not fail correctness if they are consistent
   with the gold answer and do not materially change the requested answer. Penalize
   extra details only when they contradict the gold answer, change the scope or
   conclusion, or introduce an incompatible numeric value, list item, quantity,
   or qualifier.
2. complete: The answer provides enough information to satisfy the user's probe.
   Require the final requested result and only those qualifiers or source/IFC
   caveats whose omission would likely mislead the user or change how the answer
   should be used. Do not require every list item, supporting count, document
   reference, calculation step, or caveat from the gold answer when the main
   answer is clear, actionable, and not misleading.
3. relevant: The answer directly addresses the probe question and does not answer
   the wrong project, IFC model, element, property, scope, or task.

Judge answer accuracy, not writing style or whether the answer mentions agent
trace/tool-output details.
\end{lstlisting}

\subsection{Memory Judge}

\begin{lstlisting}[style=promptstyle]
You are an expert evaluator. Your task is to judge the quality of memory retrieval and use,
not overall answer correctness. Compare the retrieved memory content against the provided
extra information. The extra information is the hidden target memory facts that
prior sessions were intended to seed. The retrieved memory content comes from the
agent trace, usually search_project_memory tool outputs.

Evaluate exactly these three dimensions:

1. retrieval_relevant: The retrieved memory content is relevant to the probe and
   the target extra information. It may include some noise, but the retrieved
   content must contain useful memory facts for the task.
2. retrieval_covers_key_facts: The retrieved memory content covers the key
   facts from the target extra information needed to answer the probe, such as source
   schedule/spec names, negative confirmations, classifications, quantities,
   units, scopes, or inclusion/exclusion rules. Exact wording is not required. retrieval_covers_key_facts must be false if retrieved_memory_content omits,
   contradicts, reverses, or replaces any central fact in target_extra_information
   needed to answer the probe.
3. answer_uses_memory: The system answer uses the retrieved memory accurately.
   It should preserve the retrieved key facts without contradiction and should
   not rely only on IFC lookup when remembered facts are needed.

Do not penalize concise answers if they correctly use the retrieved memory.

\end{lstlisting}

\end{document}